\documentclass[%
reprint,
superscriptaddress,
frontmatterverbose,
amsmath,amssymb,
pra,
]{revtex4-2}

\usepackage{amsfonts,amsmath,amssymb,amsthm}
\usepackage[normalem]{ulem}

\usepackage{graphicx}
\usepackage{dcolumn}
\usepackage{bm}

\usepackage{hyperref}

\usepackage{color}
\usepackage{amsmath}

\usepackage{lineno}
\makeatother

\newcommand{\be}{\begin{equation}}
\newcommand{\ee}{\end{equation}}

\newcommand{\br}{{\bm{r}}}

\usepackage{braket}
\usepackage{array}
\usepackage{booktabs}
\usepackage{multirow}

\usepackage{float}

\newif\ifdraft
\drafttrue
\draftfalse
\usepackage{titlesec}
\titleformat{\section}{\raggedright\bfseries\large}{\arabic{section}.}{1em}{}
\titleformat{\subsection}{\raggedright\bfseries}{\arabic{section}.\arabic{subsection}.}{1em}{}

\titlespacing*{\section}{0pt}{1em}{1pt}
\titlespacing*{\subsection}{0pt}{1em}{0pt}

\begin{document}

\title{Deep Learning Accelerated Quantum Transport Simulations in Nanoelectronics: \\From Break Junctions to Field-Effect Transistors}

\author{Jijie Zou $^{2,3}$ }
\noaffiliation 
\affiliation{School of Artificial Intelligence and Data Science, University of Science and Technology of China, Hefei 230026, China}
\affiliation{AI for Science Institute, Beijing 100080, China}
\affiliation{Centre for Nanoscale Science and Technology, Academy for Advanced Interdisciplinary Studies, Peking University, Beijing 100871, China}

\author{Zhanghao Zhouyin}
\affiliation{AI for Science Institute, Beijing 100080, China}
\affiliation{Department of Physics, McGill University, Montreal, Quebec, Canada H3A2T8}

\author{Dongying Lin}
\affiliation{Key Laboratory for the Physics and Chemistry of Nanodevices, School of Electronics, Peking University, Beijing 100871, China}

\author{Yike Huang}
\affiliation{AI for Science Institute, Beijing 100080, China}

\author{Linfeng Zhang}
\affiliation{AI for Science Institute, Beijing 100080, China}
\affiliation{DP Technology, Beijing 100080, China}

\author{Shimin Hou}
\email{smhou@pku.edu.cn}
\affiliation{Centre for Nanoscale Science and Technology, Academy for Advanced Interdisciplinary Studies, Peking University, Beijing 100871, China}
\affiliation{Key Laboratory for the Physics and Chemistry of Nanodevices, School of Electronics, Peking University, Beijing 100871, China}

\author{Qiangqiang Gu}
\email{guqq@ustc.edu.cn}
\affiliation{School of Artificial Intelligence and Data Science, University of Science and Technology of China, Hefei 230026, China}
\affiliation{AI for Science Institute, Beijing 100080, China}
\affiliation{Suzhou Institute for Advanced Research,\\ University of Science and Technology of China, Suzhou 215123, China}
\affiliation{Suzhou Big Data \& AI Research and Engineering Center, Suzhou 215123, China}

\begin{abstract}
Quantum transport simulations are essential for understanding and designing nanoelectronic devices, yet the long-standing trade-off between accuracy and computational efficiency has limited their practical applications. We present DeePTB-NEGF, an integrated framework combining deep learning tight-binding Hamiltonian prediction with non-equilibrium Green's Function methodology to enable accurate quantum transport simulations in open boundary conditions with 2-3 orders of magnitude acceleration. We demonstrate DeePTB-NEGF through two challenging applications: comprehensive break junction simulations with over $10^4$ snapshots, showing excellent agreement with experimental conductance histograms; and carbon nanotube field-effect transistors (CNT-FET) at experimental dimensions, reproducing measured transfer characteristics for a 41 nm channel CNT-FET  ($\sim 8000$ atoms, $3\times10^4$ orbitals) and predicting zero-bias transmission spectra for a 180 nm CNT ($\sim 3\times 10^4$ atoms, $10^5$ orbitals), showcasing the framework's capability for large-scale device simulations. Our systematic studies across varying geometries confirm the necessity of simulating realistic experimental structures for precise predictions. DeePTB-NEGF bridges the longstanding gap between first-principles accuracy and computational efficiency, providing a scalable tool for high-throughput and large-scale quantum transport simulations that enables previously inaccessible nanoscale device investigations.

\end{abstract}

\maketitle

\section*{Introduction}
Quantum transport simulation constitutes a cornerstone methodology for the investigation and engineering of nanoelectronic devices.  The non-equilibrium Green's function (NEGF) formalism~\cite{kadanoff1962, keldysh1965diagram, datta1997electronic}, when integrated with density functional theory (DFT)~\cite{Hohenberg1964, Kohn1965}, has emerged as the  standard method for first-principles quantum transport investigations~\cite{Taylor2001Ab,Brandbyge2002,rocha2006spin}. Nevertheless, the substantial computational overhead associated with DFT-NEGF self-consistent field (SCF) iterations imposes severe limitations on its practical applications, particularly in studying dynamic processes and large-scale systems.

This computational bottleneck is particularly acute in two critical scenarios: (i) break junction experiments~\cite{scheer2017molecular,gehring2019single}, the main platform for measuring the conductance of single
molecules, requiring statistical analysis over thousands of  configurations, 
and (ii) field-effect transistors at experimental nanoscale dimensions, where dominant quantum effects\cite{deng2023gate,iannaccone2018quantum} demand accurate first-principles modeling, but the large sizes lie beyond the reach of conventional first-principles methods. In both cases, the prohibitive computational cost restricts the studies to either selected static configurations or dramatically downsized systems, creating a disparity between computational tractability and experimental relevance. This compromise is widespread in nanoelectronics,
highlighting the urgent need for more efficient methodologies to accelerate first-principles quantum transport simulations without sacrificing accuracy.

The key to addressing this challenge is to bypass the time-consuming DFT-NEGF SCF iterations while preserving first-principles fidelity. 
Machine learning (ML) has emerged as a promising alternative to accelerate quantum transport calculations. Existing ML-based approaches include conductance prediction from atomic configurations~\cite{burkle2021deep,pan2021large,deffner2023learning,lin2025transformer}, and combining electronic and atomic features to infer transport properties~\cite{topolnicki2021combining,zhang2025physics}.
While these methods demonstrate ML's potential in transport simulations, they remain constrained to specific systems or selected transport characteristics, lacking the generalizability required for widespread adoption. A more fundamental and universal framework capable of accelerating first-principles quantum transport simulations while predicting diverse transport properties, ranging from transmission spectra to current-voltage characteristics, has yet to be established.

A more holistic solution would involve the direct prediction of the electronic Hamiltonian within the device geometries. 
By accurately modeling the fundamental electronic structure, this strategy could naturally yield all relevant transport characteristics rather than isolated properties.
Recent advances, including our recently developed deep learning tight-binding (TB) Hamiltonian method DeePTB\cite{guDeep2024,zhouyin2025learning}, along with other electronic Hamiltonian prediction techniques~\cite{liDeeplearning2022, gong2023general,zhongTransferable2023a}, have demonstrated impressive accuracy in predicting electronic structures.
However, these approaches are limited to systems with periodic or isolated boundary conditions, leaving the unique challenges of open boundary conditions in device modeling unaddressed.

In this work, we introduce DeePTB-NEGF, a transformative framework extending the capabilities of electronic structure prediction to open boundary systems, a critical requirement for quantum transport in realistic devices with terminals. By achieving a 2 to 3 orders of magnitude acceleration compared to conventional DFT-NEGF approach while maintaining first-principles accuracy, DeePTB-NEGF enables quantum transport simulations of complete experimental devices at previously inaccessible scales. The versatility and accuracy of DeePTB-NEGF are demonstrated in two challenging and representative applications.

First,we apply DeePTB-NEGF to break junction systems, demonstrating its capability for high-throughput simulations of dynamic processes. By modeling the complete elongation process (over $10^4$ snapshots) of metallic contacts and molecule junctions, DeePTB-NEGF overcomes the immense computing cost of traditional DFT-NEGF while achieving excellent agreement with   experimental data~\cite{guo2024effectively}. This application highlights DeePTB-NEGF's ability to tackle statistical challenges requiring extensive simulations across numerous configurations.

Second, we demonstrate DeePTB-NEGF's capacity to handle complex
gate-controlled devices through simulations of carbon nanotube field-effect transistors (CNT-FETs) with local bottom gate. 
Unlike two-terminal transport calculations, these simulations incorporate the bias among source, drain, and gate, necessitating self-consistent solutions of  Poisson and quantum transport equations to properly capture gating effects. 
Our method accurately reproduces transfer characteristics for CNT-FETs with channel lengths up to 41 nm ($\sim$ 8000 atoms), showing excellent consistency with experimental measurements~\cite{franklin2012sub}. Furthermore, we successfully predict the transmission spectrum of a 180 nm CNT ($\sim$$3\times 10^4$ atoms and ~$10^5$ orbitals), demonstrating DeePTB-NEGF's potential for large-scale device simulations with first-principles accuracy.
 
These results establish DeePTB-NEGF as a transformative framework that finally bridges the long-standing divide between theoretical modeling and experimental reality in quantum transport simulations. By enabling accurate, efficient modeling of device geometries at realistic scales, our method opens new horizons for the design and understanding of next-generation devices.

\section*{Results}
\subsection*{Theoretical framework and workflow}
\begin{figure}[tbp!]
    \includegraphics[width=8.0 cm]{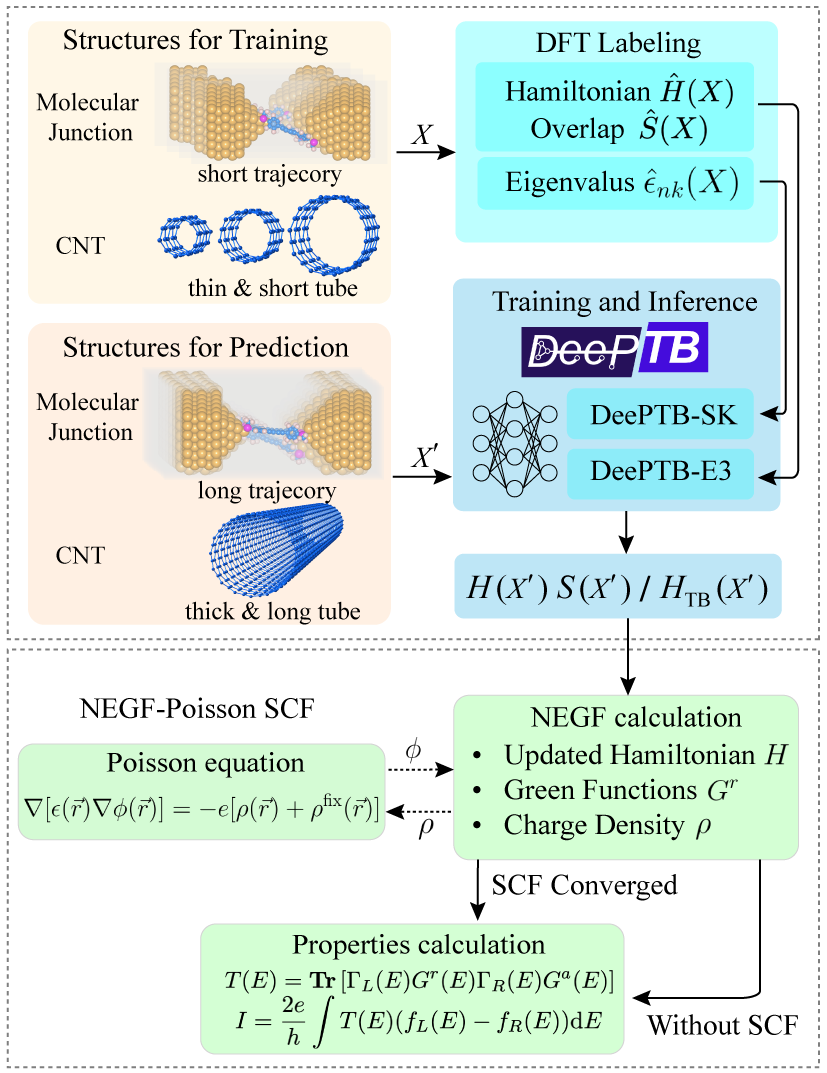}
    \caption{Schematic illustration of DeePTB-NEGF workflow. Atomic configurations $X$ and $X'$ are structures for training and prediction respectively. Reference Hamiltonians/overlap matrices $\hat{\bm{H}}(X)$/$\hat{\bm{S}}(X)$ and reference eigenvalues $\hat{\epsilon}_\text{nk}(X)$ are obtained from first-principles calculations, while $\bm{H}(X')$/$\bm{S}(X')$ and $\bm{H}_\text{TB}(X')$ represent predictions from DeePTB model with E3 method and SK method respectively. $T(E)$ denotes the transmission coefficient at energy $E$ and $I$ is the current through the device.}
    \label{fig:workflow}
\end{figure}

Efficient first-principles quantum transport simulations require accurate electronic structure description while avoiding the computationally expensive DFT-NEGF SCF iterations. As illustrated in Fig.~\ref{fig:workflow}, our DeePTB-NEGF framework integrates deep learning-based Hamiltonian prediction and the NEGF formalism to enable high-throughput and large-scale quantum transport calculations. Below we elaborate on each component of this integrated framework.

\bigskip
\noindent\textbf{A. Hamiltonain parameterization} 
DeePTB offers two complementary strategies for constructing electronic Hamiltonians: (1) the environment-dependent Slater-Koster(SK) TB Hamiltonians~\cite{guDeep2024} and (2) Kohn-Sham (KS) Hamiltonians in the linear combination of atomic orbitals (LCAO) basis with E(3) equivariant graph neural networks\cite{zhouyin2025learning}. These strategies represent different trade-offs between computational efficiency and physical accuracy.

For the SK TB Hamiltonians, DeePTB trains local environment-dependent SK integrals  
 $h^{\text{env}}_{ll^\prime{\zeta}}$ as\cite{guDeep2024}:
\begin{equation}
    h^{\text{env}}_{ll^\prime{\zeta}} =  h_{ll^\prime{\zeta}}(r_{ij}) \times \left[1+\Phi_{ll^\prime\zeta}^{o_i,o_j}\left(r_{ij},\mathcal{D}^{ij}\right)\right]	
    \label{eq:eqsk}
\end{equation}
where $h_{ll^\prime{\zeta}}(r_{ij})$ is the conventional SK integral, $\Phi_{ll^\prime\zeta}^{o_i,o_j}$ introduces neural network-based environmental corrections through the descriptor $\mathcal{D}^{ij}$.This descriptor captures the chemical environment surrounding the atoms $i$ and $j$. $l,l^\prime$ denote atomic orbitals, $\zeta$ the bond types, and $r_{ij}$ the interatomic distance.   
By training on the DFT eigenvalues $\hat{\epsilon}_{n\bm{k}}(X)$, the SK integrals as functions of the local environment can be predicted for unseen structures.

For the LCAO KS Hamiltonians, DeePTB predicts multiple quantum operators (Hamiltonian, overlap and density matrices) through E(3) equivariant strategy~\cite{zhouyin2025learning}, preserving rotational symmetry through angular momentum coupling:
\begin{equation}
O^{i,j}_{l_1,l_2,m_1,m_2}=\sum_{l_3,m_3}C_{(l_1,m_1)(l_2,m_2)}^{(l_3,m_3)}o_{l_3,m_3}^{i,j}
\label{eqe3}
\end{equation}
where $O$ represents the quantum operators, $C_{(l_1,m_1)(l_2,m_2)}^{(l_3,m_3)}$ are the Clebsch-Gordan coefficients and $o_{l_3,m_3}^{i,j}$ are the output node or edge features from the equivariant graph neural networks.
After training on converged Hamiltonians and overlap matrices $\hat{\bm{H}}(X)$ and $\hat{\bm{S}}(X)$ obtained from DFT-NEGF, the features $o_{l_3,m_3}^{i,j}$ would be determined.
Both strategies provide efficient access to electronic Hamiltonians while maintaining first-principles accuracy, though based on different theoretical frameworks. 

\bigskip
\noindent\textbf{B. Open Boundary Conditions Treatment} 
Accurate quantum transport simulations require proper treatment of open boundary conditions to correctly model electron injection and extraction at the electrode-device interfaces. 
In the NEGF formalism, these open boundary conditions are incorporated through self-energies $\bm{\Sigma}$ that describe the coupling between the central scattering region and the semi-infinite electrodes with unit cells identical to the bulk structure\cite{datta1997electronic}. Conventional DFT-NEGF approaches usually calculate electrode Hamiltonians from separate bulk simulations, then use these to construct self-energies with appropriate Fermi level alignment with the scattering region\cite{rocha2006spin,Brandbyge2002,papior2017improvements}.

\begin{figure}[tbp!]
    \includegraphics[width=7.5 cm]{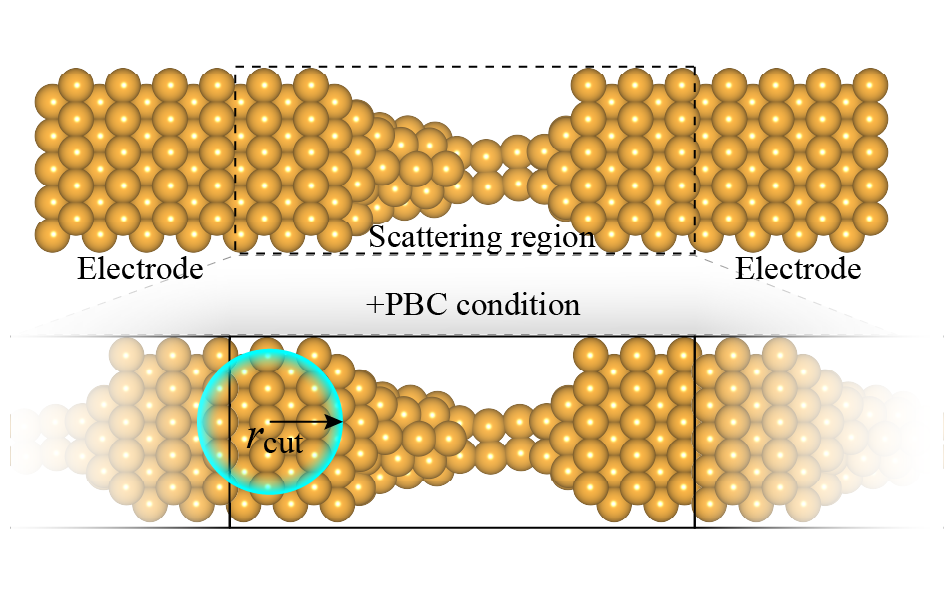}
\caption{Illustration of device with open boundary condition (gold atomic contact as an example) and DeePTB-NEGF treatment. Top panel: configuration for transport calculation consisting of a central scattering region connected to two semi-infinite electrodes. Bottom panel:configuration for DeePTB training with periodic boundary conditions. The local environment inside the cutoff radius $r_\text{cut}$ (marked by bubble) for the atoms near the edge of electrode extension is bulk-like. }
\label{fig:SR_PBC}
\end{figure}

For deep learning models to work effectively with open boundary conditions, they must accurately predict the electronic structure not only in the scattering region but also in the semi-infinite bulk  electrodes, as depicted in the upper pannel of  Fig.~\ref{fig:SR_PBC} . However, we discovered that simply training on primitive unit cells fails to produce correct Hamiltonians for electrode extensions, even with identical local environments. These  Hamiltonians differ from true bulk by a constant energy shift due to long-range effects from the scattering region, creating a transferability problem.

We address this by introducing a constant shift freedom in the Hamiltonian representation:
\begin{equation}
   H^\prime = H + \lambda \cdot S 
\end{equation}
where $\lambda$ is determined during training by comparing DeePTB-predicted and DFT-calculated Hamiltonian blocks. This allows the training dataset to include both primitive unit cells and scattering regions simultaneously.
To ensure complete electrode extension description, we
train on the entire scattering region with electrode extensions to create bulk-like environments within the cutoff radius $r_\text{cut}$ for atoms at the edge, as shown in the lower panel of Fig.~\ref{fig:SR_PBC}.
This leverages DeePTB's locality features—either through the localized equivariant message-passing in DeePTB-E3\cite{zhouyin2025learning} or the intrinsic locality of tight-binding in DeePTB-SK\cite{guDeep2024}. This locality captures relevant information nearly strictly within $r_\text{cut}$, minimizing unnecessary long-range connections. Furthermore, using periodic boundary conditions in training (lower panel, Fig.~\ref{fig:SR_PBC}) reduces the required electrode extension layers, improving the training efficiency.

\bigskip
\noindent\textbf{C. NEGF formalism} 
With the Hamiltonians efficiently predicted by DeePTB, quantum transport properties are calculated using the NEGF formalism. The NEGF approach provides a rigorous framework for quantum transport by incorporating open boundary conditions through self-energies $\bm{\Sigma}$ that describe the coupling between the device and semi-infinite electrodes. For a two-terminal system, the transmission spectrum $T(E)$, quantifying the energy-dependent electron transmission probability between electrodes, is given by:
\begin{equation}
    T(E) = \textbf{Tr} \left[\Gamma_L(E) G^r(E) \Gamma_R(E) G^a(E)\right]
\end{equation}
where $\Gamma_{L/R} = i(\bm{\Sigma}_{L/R}^r-\bm{\Sigma}_{L/R}^a)$ is the broadening function for the left ($L$) and right ($R$) electrodes and $G^{r/a}(E)=[(E \pm i\eta)S-H-\bm{\Sigma}_L^{r/a}-\bm{\Sigma}_R^{r/a}]^{-1}$ is the retarded/advanced Green's function with $\eta$ being a positive infinitesimal and $H/S$ being the Hamiltonian/overlap matrices. 

The current ($I$) and zero-bias conductance ($G$) at the zero-temperature limit are then given by~\cite{scheer2017molecular}:
\begin{align}
    I = \frac{2e}{h} & \int T(E) (f_L(E)-f_R(E)) \text{d}E\\
    G & = \frac{2e^2}{h} T(E_f) = T(E_f)\mathrm{G}_0
\end{align}
where $f_{L/R}(E)$ are Fermi-Dirac distributions corresponding to the Fermi level of the left and right electrodes, $E_f$ is the Fermi energy, $\mathrm{G}_0 = 2e^2/h$ is the quantum conductance, $e$ is the elementary charge, and $h$ is the Planck's constant.With Hamiltonians predicted by DeePTB, we further apply Bloch's theorem to calculate the self-energy~\cite{papior2017improvements}and utilize recursive Green's function method~\cite{anantram2008modeling} with greedy algorithm~\cite{Pecchia_2008_NEGF} to substantially enhance efficiency.


\bigskip
\noindent\textbf{D. NEGF-Poisson Self-Consistency} 
To model realistic electronic devices operating under finite bias, it is essential to incorporate electrostatic effects through self-consistent coupling between quantum transport and electrostatics. In the DeePTB-NEGF framework, this is achieved through the NEGF-Poisson SCF procedure, which is particularly important for simulating field-effect transistors with gate control.

This self-consistency is achieved through the Gummel iteration scheme~\cite{gummel1964self}, which begins with the equilibrium TB Hamiltonian $H_0$ and iteratively solves for the non-equilibrium potential profile using the Poisson equation~\cite{fiori2006three,ahn2022fully}
\begin{equation}
\nabla[\epsilon (\vec{r})\nabla \phi_\text{n+1}(\vec{r})] = -e[\rho_\text{n}(\vec{r})+\rho^{\text{fix}}(\vec{r})]
\label{eq:poi}
\end{equation}
with the Newton-Raphson method on a discrete real-space grid. Here $\epsilon$ is the dielectric constant, $\phi_\text{n+1}$ and $\rho_\text{n}$ are the electrostatic potential and free charge number concentration in the $n$-th iteration. The term $\rho^{\text{fix}}$ represents the average equivalent charge associated with ionized donors or acceptors at each atomic site within the doped region.

In practical calculations, $\rho_\text{n}$ is obtained as Mulliken charge at all atomic sites based on the Hamiltonian $H_\text{n}$ from the n-th iteration. The boundary conditions for the Poisson equation include constraints from gate voltage $V_\text{gs}$ and drain-source bias $V_\text{ds}$. Within DeePTB-NEGF framework, the predicted TB Hamiltonian is updated according to the electrostatic potential as:
\begin{equation}
    H_{n}^i = H_0^i - e\phi_\text{n}(\br_i)
\end{equation}
where $H_{n}^i$ and $\phi_\text{n}(\br_i)$ are the $n$-th iteration Hamiltonian onsite block and electrostatic potential at site $i$, respectively. This NEGF-Poisson procedure enables the self-consistent treatment of quantum transport and electrostatics, essential for simulating realistic device behavior under bias conditions.

\bigskip
Having established the theoretical foundation and computational framework of DeePTB-NEGF, we now demonstrate its practical capabilities through two representative applications. These applications highlight the framework's ability to address long-standing challenges in quantum transport simulation: statistical analysis of break junction experiments and realistic simulation of gate-controlled devices. The first application demonstrates how DeePTB-NEGF enables high-throughput calculations for statistical analysis, while the second showcases its capacity to handle large-scale systems with realistic dimensions.
\subsection*{Application 1: Break Junction Systems}
Break junction experiments, primarily implemented through mechanically controllable break junctions or scanning tunneling microscopy break junctions~\cite{scheer2017molecular,gehring2019single}, provide a powerful platform for investigating quantum transport at the atomic and molecular scales. During a typical break junction elongation process, configurations evolve from metallic contacts with quantized conductance to the molecular junctions bridging nanoscale gaps~\cite{xu2003measurement}. 
Statistical nature of these experiments requires analyzing thousands of conductance measurements across multiple breaking events - a task prohibitively expensive for conventional first-principles calculations. 
The DeePTB-NEGF framework developed here enables efficient simulation of the break junction process while maintaining first-principles accuracy. Below we demonstrate its capabilities in both metallic contacts and molecular junctions, providing direct comparison with experimental measurements and atomic-level insights into the quantum transport mechanisms.

\begin{figure*}[htbp!]
\includegraphics[width=16 cm]{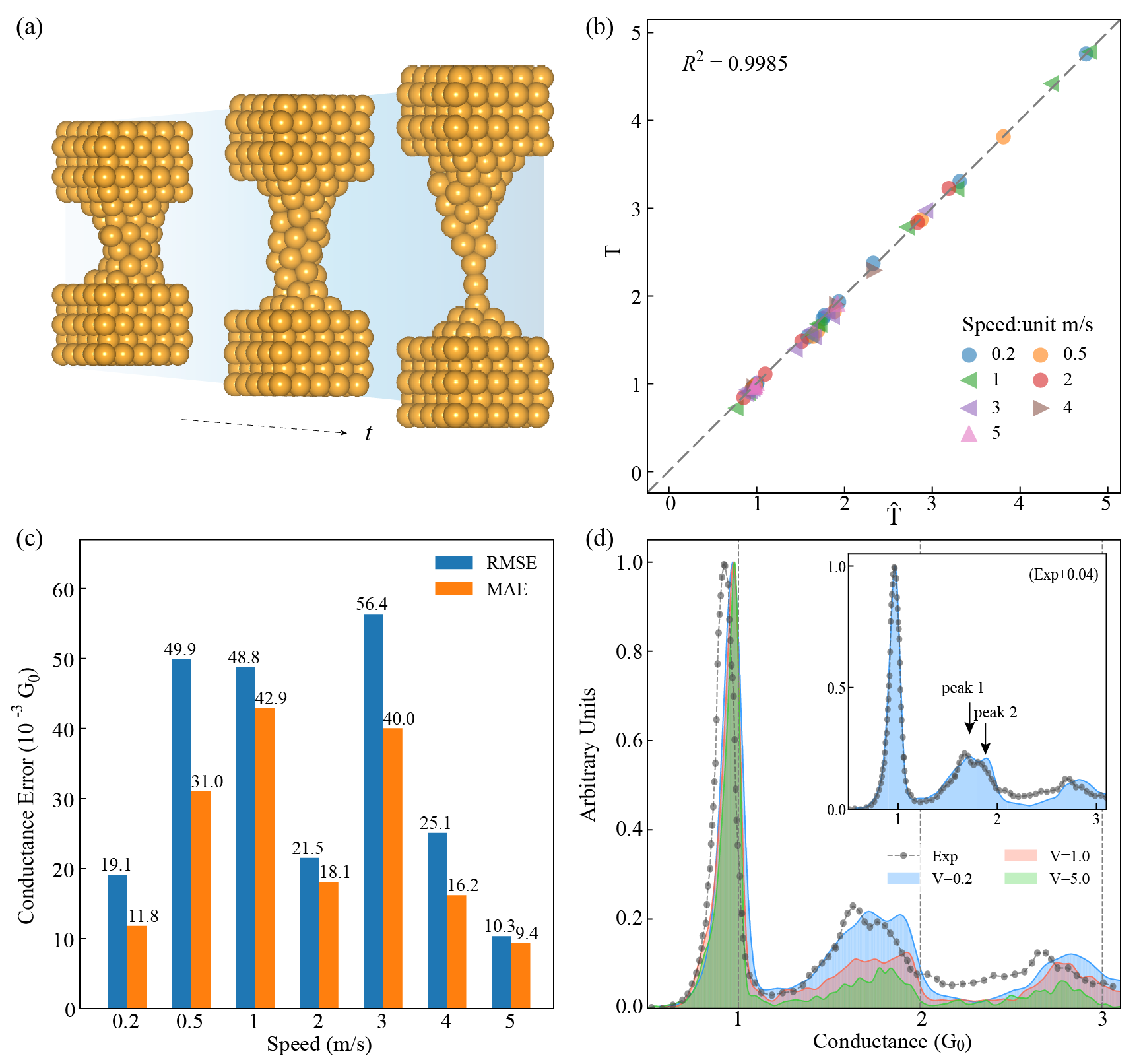}
\caption{DeePTB-NEGF simulation results for gold contacts. (a) Three representative snapshots in one breaking junction process. (b) Comparison for transmission at the Fermi level between the DeePTB-NEGF method ($T$) and DFT-NEGF calculations($\hat{T}$) in structures obtained with different elongation speeds. (c) RMSE and MAE for zero-bias conductance between DeePTB-NEGF and DFT-NEGF results. Both RMSE and MAE are below $6\times10^{-2} \text{G}_0$, ensuring sufficient accuracy for reliable conductance statistical analysis. (d) Conductance histograms from 10,119 configurations sampled from 85 elongation processes at three different speeds ($v=0.2$, $1.0$, and $5.0\text{ m/s}$) compared with experimental measurements. Inset: experimental $1\mathrm{G}_0$ peak aligns with the $v=0.2$ m/s case after shifting the experimental histogram by $+0.04 \mathrm{G}_0$.}
\label{fig:Au_bj}
\end{figure*}

\bigskip
\noindent\textbf{A. Metallic contacts}
Gold atomic contacts formed in break junction experiments (shown in Fig.~\ref{fig:Au_bj}(a)) exhibit distinct quantum transport features including conductance quantization and ballistic transport\cite{yu2016atomic}. 
As electrodes separate, the conductance evolves through characteristic plateaus, corresponding to specific atomic-scale contact configurations\cite{rego2003role}.
It requires over thousands of breaking events to understand these quantum transport features through statistical analysis, making theoretical reproduction of conductance histograms at the first-principles level computationally challenging.

To simulate quantum transport through gold contacts, we construct a junction containing 304 gold atoms arranged along the [100] direction, with two bulk-like electrode extensions (100 atoms each in $5\times 5 \times 4$ supercells) shown as the scattering region in Fig.~\ref{fig:SR_PBC}.
Break junctions experiments were simulated through molecular dynamics simulations at 150K using the machine learning potential~\cite{andolina2021robust} implemented in DeePMD~\cite{zhangDeep2018}, with the simulation box being the scattering region under periodic boundary conditions. 
During elongation, the electrode sections are fixed as bulk structure and move in opposite directions at speeds ranging from 0.2 to 5.0 m/s on each side. The resulting structures exhibit characteristic bipyramidal shapes consistent with experimental high-resolution transmission electron microscopy (TEM) observations~\cite{Lagos_2010}.

To construct the DeePTB model for gold contacts, we randomly selected 122 configurations from 4 independent trajectories at $v=\pm5.0$ m/s labeled with Hamiltonians and overlaps using DFT-NEGF. Details of dataset and training settings can be found in Methods.
Building on DeePTB-NEGF framework's strategy in open-boundary systems, the model accurately predicts the electronic Hamiltonian for the complete atomic contact system (top pannel of Fig.~\ref{fig:SR_PBC}), achieving RMSE of $7.57\times 10^{-4}$ eV for Hamiltonians, and $4.73\times 10^{-5}$ for overlap matrices on the validation set.
More importantly, despite being trained only on configurations from high-speed trajectories ($v = 5.0$ m/s), the model demonstrates outstanding transferability across different elongation speeds. For 55 randomly selected configurations spanning speeds from 0.2 m/s to 5.0 m/s, the predicted zero-bias conductance achieves exceptional agreement with DFT-NEGF results ($R^2 = 0.9985$) as shown in Fig.\ref{fig:Au_bj}(b), with both RMSE and MAE below $60\times 10^{-3} \mathrm{G}_0$ across all elongation speeds, as illustrated in Fig.\ref{fig:Au_bj}(c). This accuracy ensures sufficient accuracy for reliable conductance statistical analysis.

With the model validated, we then demonstrate its efficiency in high-throughput calculations by analyzing  10,119 configurations randomly sampled from 85 elongation processes at speeds of 0.2, 1.0, and 5.0 m/s - a scale previously inaccessible to first-principles methods.
The calculations were accelerated using the Bloch theorem for self-energy evaluation and exploiting the tridiagonal block structure of Hamiltonian matrices. 
The predicted conductance histograms, normalized w.r.t. their respective $1\mathrm{G}_0$ peaks, are presented in Fig.~\ref{fig:Au_bj}(d). The prominent peak at $1\mathrm{G}_0$ corresponds to monatomic chains, a well-established signature in gold metallic break junctions. 
Different elongation speeds lead to distinct variations in conductance peak shapes, particularly in the range from $1\mathrm{G}_0$ to $2\mathrm{G}_0$. This variation arises from the suppression of collective atomic relaxations at higher elongation speeds, which prolongs the stability of monatomic chains and therefore increases the height of $1\mathrm{G}_0$ peak before breaking\cite{burkle2021deep,Sutton1996Force}. Consequently, the relative intensity of conductance features beyond $1\mathrm{G}_0$ is systematically reduced in the normalized histograms.
Note that the simulation speeds (0.2-5.0 m/s) are necessarily higher than experimental rates ($10$ pm/s to $100$ nm/s)\cite{scheer2017molecular}, because the experimental breaking processes would require prohibitively large MD simulation times\cite{Lagos_2010}.
Nevertheless, even at these elevated speeds, our simulated conductance histogram at $v=0.2$ m/s shows good agreement with experimental measurements, despite a minor shift (0.04$\mathrm{G}_0$) in peak positions.
This minor shift may result from structural perturbations or defects in the electrode, or from additional scattering processes in experiments, which are ignored in the simulations.
By shifting the experimental histogram, we achieve excellent shape alignment of the $1\mathrm{G}_0$ peaks as shown in the inset of Fig.~\ref{fig:Au_bj}(d). The aligned histograms reveal remarkable agreement between calculation and experiment~\cite{Lagos_2010}, particularly in reproducing two characteristic conductance peaks between $1\mathrm{G}_0$ and $2\mathrm{G}_0$.

These results demonstrate that our DeePTB-NEGF framework can efficiently and accurately predict the quantum transport properties of gold contacts, capturing the statistical features of the conductance histograms consistent with the experimentally observed characteristics.


\begin{figure*}[htbp!]
\includegraphics[width=16 cm]{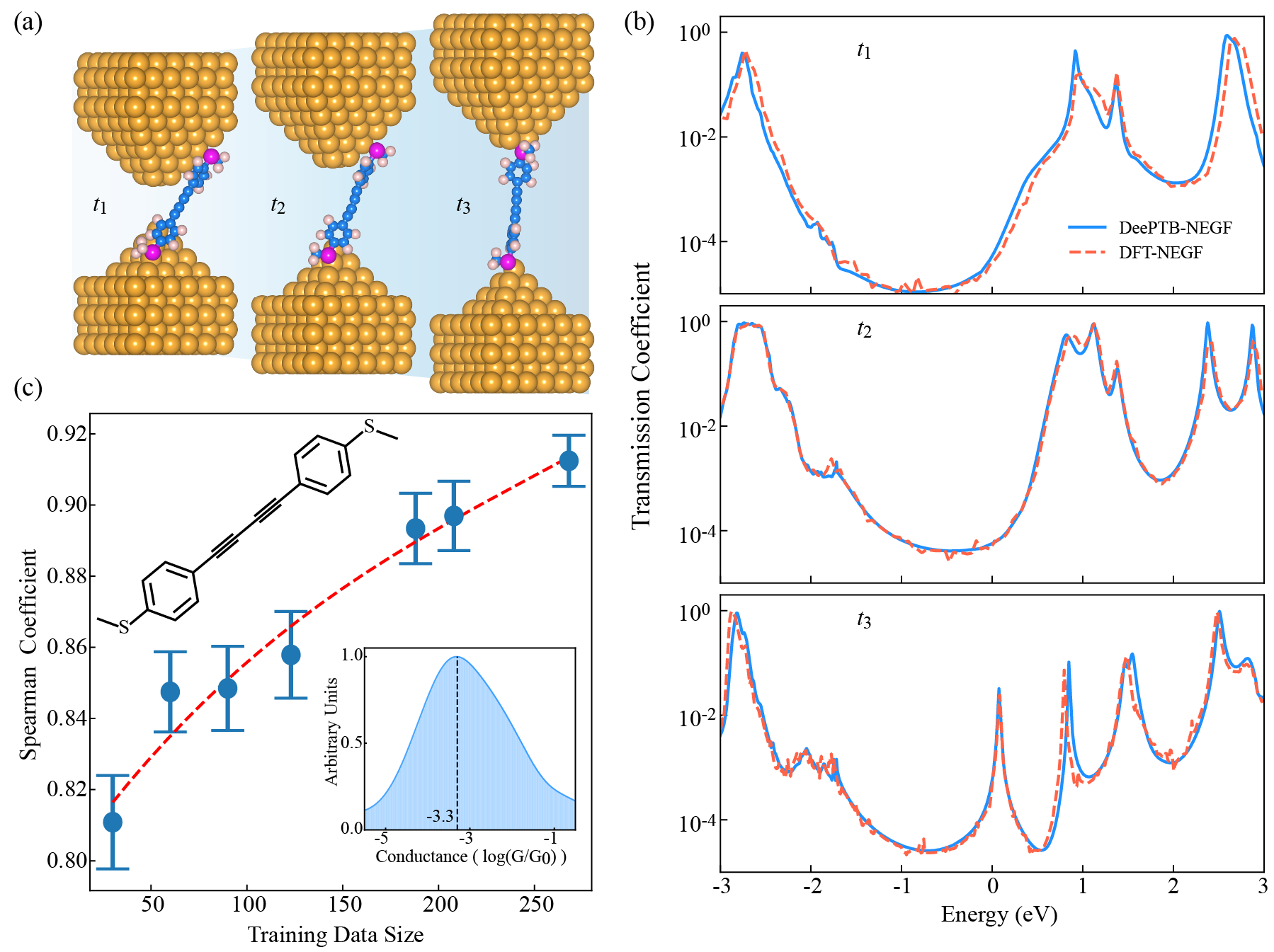}
\caption{DeePTB-NEGF simulation results for molecular junctions. (a) Three sequential snapshots of the molecular junction breaking process at equal time intervals ($t_1$, $t_2$, and $t_3$). (b) Transmission spectra corresponding to the three snapshots in (a), and Fermi energy set to $0$. (c) Spearman correlation coefficient between DeePTB-NEGF and DFT-NEGF transmission spectra within the energy range of $-0.5$ to $0.5$ eV as a function of training dataset size. Blue dots show mean values with standard deviation error bars. Top left inset: Structure of the $\pi$-conjugated molecule with sulfur methyl anchoring groups. Bottom right inset:Conductance histogram from 590 configurations across 11 stretching trajectories, showing a prominent peak at $10^{-3.3} \mathrm{G}_0$.}
\label{fig:Aumol}
\end{figure*} 
\bigskip
\noindent\textbf{B. Single-molecule junctions} 
Unike metallic contacts, single-molecule junctions present additional challenges for quantum transport calculations due to their complex chemical environments and dynamic metal-molecule interfaces. Their transmission spectra are particularly sensitive to molecular conformations and contact geometries, requiring extensive sampling to capture the dynamic features during junction formation and breaking that has been prohibitively expensive with traditional DFT-NEGF methods.
Consequently, previous first-principles simulation studies were typically limited to a few selected relaxed structures, missing crucial dynamic aspects of transport during junction evolution. Our DeePTB-NEGF framework overcomes this limitation by enabling high-throughput transmission spectra calculations throughout the complete junction formation and breaking process.

We investigated a classical $\pi$-conjugated system with thiomethyl group as anchoring groups (top left inset of Fig.~\ref{fig:Aumol}(c)), chosen for its rich chemical environments and flexible configurations due to three types of carbon bonds and the dihedral angle between the two benzene rings. 
The scattering region consists of a total of 478 atoms, with the molecule connected to gold electrodes with a bi-pyramidal shape, where each electrode principal layer contains 144 atoms ($6\times 6 \times 4$ supercell). During stretching, the gold electrodes move in opposite directions at a speed of $v=\pm 2$ m/s, with three representative snapshots at $t_1$, $t_2$, $t_3$ in equal time intervals shown in Fig.~\ref{fig:Aumol}(a).
To construct the DeePTB model for molecular junctions, we randomly sampled 268 configurations from 14 breaking junction trajectories as the training set labeled with electronic Hamiltonians and overlaps using DFT-NEGF. More details of dataset and training settings can be found in Methods.

To optimize model performance, we systematically studied the impact of training set size by training models using varying numbers of configurations (30, 60, 90, 123, 188, 208, 268) randomly sampled from the 14 breaking junction trajectories. The prediction accuracy was evaluated using the Spearman coefficient~\cite{spearman1987proof} between DeePTB-NEGF and DFT-NEGF transmission spectra in the energy range of -0.5 to 0.5 eV, which measures how well two variables can be described by a monotonic function. 
The coefficients are evaluated utilizing a validation set of 30 configurations from a different trajectory to test out-of-distribution generalization.
As shown in Fig.~\ref{fig:Aumol}(c), both the mean and standard deviation of the correlation coefficient improve with increasing training set size, reaching a high value of 0.912 with 268 configurations. The diminishing rate of improvement suggests that the model performance is approaching convergence.
 

To further validate the statistical assessment of model accuracy indicated by the Spearman correlation coefficient of 0.912, we examine the performance of our model trained with 268 configurations by comparing transmission spectra for three representative configurations captured at times $t_1$, $t_2$, and $t_3$ during the breaking junction process shown in Fig.~\ref{fig:Aumol}(a). 
As illustrated in Fig.~\ref{fig:Aumol}(b), the DeePTB-NEGF predicted transmission spectra closely match the DFT-NEGF results across different junction configurations, accurately reproducing both the positions and shapes of transmission peaks. Transmission comparisons in more snapshots are presented in Supplementary materials(SM) Fig.~S1.
Analysis of these transmission spectra further reveals distinct behavior of frontier molecular orbital contributions: the HOMO-dominated transmission peak maintains its energetic position relative to the Fermi level, while the LUMO-dominated peak shifts toward the Fermi level during junction elongation. This characteristic evolution of transmission peaks indicates that the conductance is LUMO-dominated,  which is consistent with previous experimental and theoretical studies\cite{guo2024effectively}.
Furthermore, to enable statistical comparison with experimental measurements, we calculated the zero-bias conductance for 590 configurations sampled from 11 breaking junction trajectories. The resulting conductance histogram (bottom right inset of Fig.~\ref{fig:Aumol}(c)) shows a pronounced peak at $10^{-3.3}\mathrm{G}_0$, in good agreement with the experimental value of $10^{-3.6}\mathrm{G}_0$\cite{guo2024effectively}.

To quantify the computational advantage of our approach, we benchmarked DeePTB-NEGF against conventional DFT-NEGF using a molecular junction snapshot on moderate computing hardware (a 28-core CPU). By systematically scaling the system size up to 4,798 atoms, we observed that DeePTB-NEGF achieves a remarkable $2 - 3$ orders of magnitude speedup, as detailed in Fig.~S2. The largest system's transmission near the Fermi level was calculated in just 633 seconds, compared to projected $10^5$ seconds (over 27 hours) for DFT-NEGF based on the linear extrapolation. 
More details can be found in SM Sec.
~S2. This dramatic efficiency improvement transforms what were previously prohibitive calculations into routine tasks that can be performed on modest hardware.

From these investigations of break junction systems, we have demonstrated the applicability and accuracy of the DeePTB-NEGF approach in simulating quantum transport properties in the two cases of metallic atomic contacts and single-molecule junctions. Our method successfully captures the key features of the conductance evolution during the junction-breaking process and reproduces the main characteristics of the experimental conductance histograms. The excellent agreement between the DeePTB-NEGF and DFT-NEGF results further validates its accuracy in describing the electronic structure of the junctions, while our efficiency benchmarks reveal a remarkable $2 - 3$ orders of magnitude speedup over conventional calculations.
These results establish DeePTB-NEGF as a transformative method for studying quantum transport in break junction systems, finally enabling the statistical analysis necessary for meaningful comparison with experimental measurements while maintaining first-principles accuracy.

\bigskip
\subsection*{Applicaiton 2: Carbon Nanotube Field-effect Transistor}\label{sec:cnt}

\begin{figure*}[btp!]
    \includegraphics[width=16 cm]{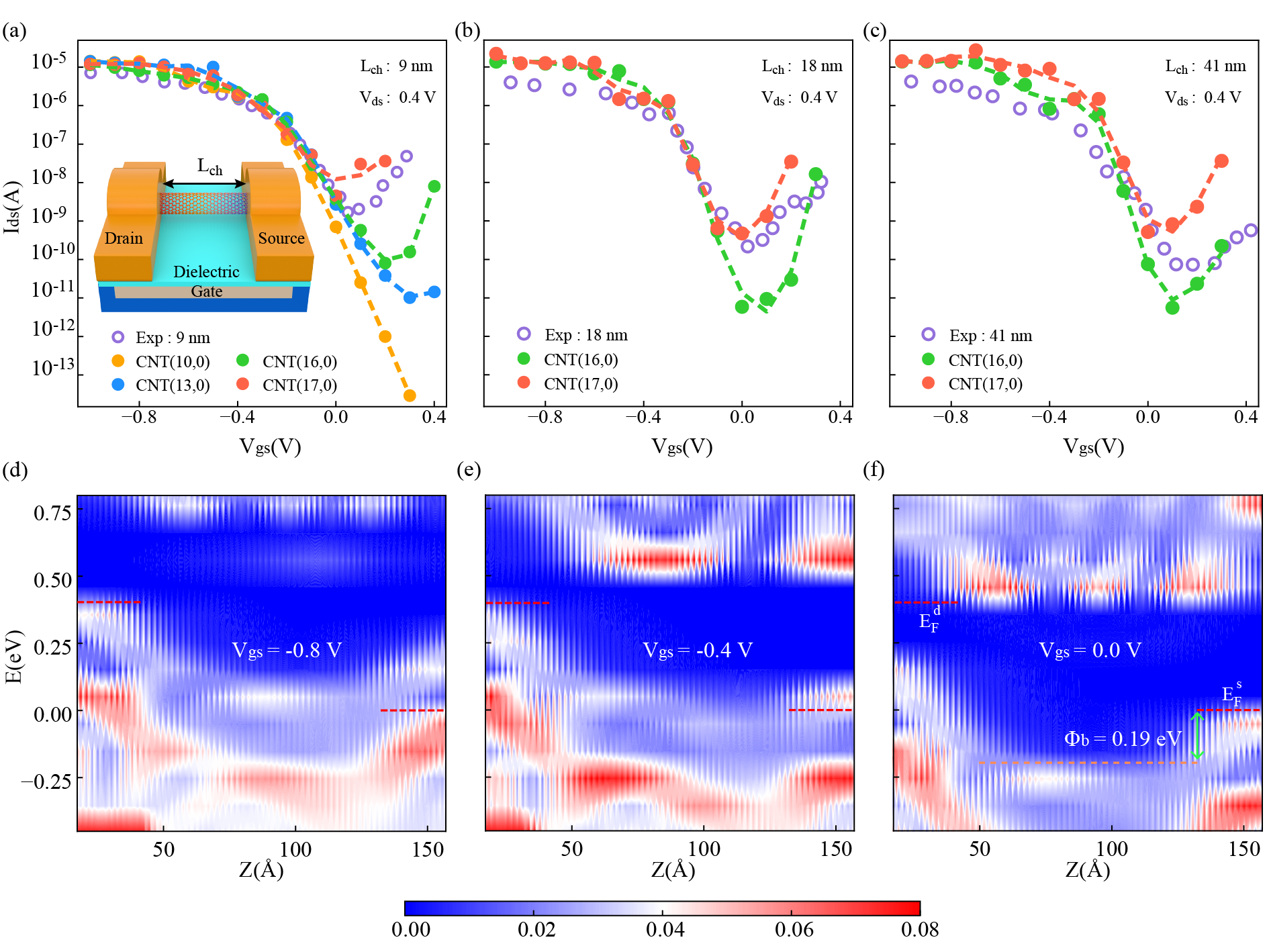}
    \caption{DeePTB-NEGF-Poisson SCF simulation for local bottom gate carbon nanotube (LBG) field-effect transistor. (a)Transfer characteristics at drain-source bias $V_\text{ds}=0.4$ V and dope concentrations as  $8.89\times10^{8} \text{m}^{-1}$ for LBG CNT FETs using different CNT diameters. The channel length $L_\text{ch}$ for all FETs are 9nm. Inset: Schematic of the transistor geometry, where $L_\text{ch}$ denotes channel length. (b)-(c) Transfer characteristics at drain-source bias $V_\text{ds}=0.4$ V and dope concentrations as  $8.89\times10^{8} \text{m}^{-1}$ for LBG CNT FETs with channel length $L_\text{ch}$ as 18 nm and 41 nm. (d)-(f) Position-resolved local density of states for CNT$(17,0)$ along the transport direction($z$ axis) with gate voltage $V_\text{gs}=-0.8 \text{V}, -0.4 \text{V}, 0.0 \text{V}$.  The left and right electrodes are drain and source respectively. The red dashed lines indicate the Fermi level for source and drain. The orange dashed line in $V_\text{gs}=0.0$ case indicates the top of valence band inducing a barrier $\Phi_B=0.19$ eV for holes.}
    \label{fig:CNT_all}
\end{figure*}

Having demonstrated the capability of DeePTB-NEGF in break junction systems, we now turn to  carbon nanotube  field-effect transistors (CNT-FET) that present substantially  challenges in scale and complexity. CNT with their quasi-one-dimensional structure and long mean free path, enable ballistic transport at nanoscale dimensions, making them promising candidates for high-performance electronic devices~\cite{iijima1991helical,franklin2012sub,peng2014carbon,pitner2020sub,cao2023future}. Previous theoretical studies have typically relied on empirically parameterized Hamiltonians \cite{alam2005performance, guo2004numerical,leonard2006properties,franklin2012sub} or employed tubes much smaller than experimental dimensions \cite{xu2021can}. 
These compromises have created a persistent gap between theoretical simulations and experimental reality.

DeePTB-NEGF overcomes these limitations by enabling first-principles quantum transport simulations at experimental scales. To validate its computational capabilities, we first applied our method to predict transmission spectra for CNTs of increasing length. As shown in SM Sec.~S4, with DeePTB-SK, our approach accurately reproduces DFT-NEGF results for short tube (9 nm) while scaling efficiently to CNTs with length up to 180 nm—containing approximately 30,000 atoms and $10^5$ orbitals. This computational advantage demonstrates the potential of DeePTB-NEGF for large-scale device simulations.

Building on this foundation, we applied our framework to the more challenging and practically relevant case of CNT-FETs with gate control. 
The critical distinction in FETs is the presence of gate-modulated electrostatics, which necessitates a self-consistent solution of the Poisson equation coupled with the NEGF formalism~\cite{sanchez2021top,guo2004performance}. As illustrated in Fig.~\ref{fig:workflow}, an integrated NEGF-Poisson SCF procedure is implemented in the DeePTB-NEGF formalism, enabling the quantum transport simulation under finite bias conditions. For these simulations, we employed the environment-dependent SK TB Hamiltonians predicted by DeePTB (Eq. \ref{eq:eqsk}), which offer advantages over Kohn-Sham Hamiltonian models due to their smaller, sparser matrices and direct compatibility with the NEGF-Poisson framework. The complete NEGF-Poisson procedure is detailed in SM Sec.~S5, and our implementation has been validated against the established TB-NEGF code NanoTCAD ViDES~\cite{fiori2006three,marian2023multi}, as described in SM Sec.~S6.

We now apply the DeePTB-NEGF framework to study the transport properties of a local bottom gate (LBG) CNT-FET with experimental dimensions, demonstrating the capability of our method to simulate realistic nanoelectronic devices. 
The LBG CNT-FET geometry, shown in the inset of Fig.~\ref{fig:CNT_all}(a), 
has been experimentally demonstrated to scale from 41 nm to 9 nm channel lengths without short-channel effects.\cite{franklin2010length,franklin2012sub}
By simulating this specific device with different channel lengths, we aim to validate our method against experimental results and explore the scaling limits of CNT-FETs.

Firstly we investigate the influence of CNT diameter on device performance, we simulated LBG CNT-FETs (channel length $L_\text{ch}=9$ nm) with various chirality  indices $(10,0),(13,0)$, $(16,0)$ and $(17,0)$, corresponding to diameters ranging from 0.79 nm to 1.35 nm. This range includes the 1.3 nm diameter CNTs used in the experiment reported in Ref.\cite{franklin2012sub}, with CNT(16,0) and CNT(17,0) being the most comparable at diameters of 1.27 nm and 1.35 nm, respectively.
We trained a $sp$-DeePTB model with environment-dependent SK parameter using the DFT eigenvalues of the CNT$(7,0),(10,0),(13,0)$ and $(16,0)$ structures as the training data set, allowing us to predict the TB Hamiltonian for CNT$(17,0)$. 
A comparison of band structures between DFT and DeePTB is illustrated in Fig.~S3 in SM, validating the accuracy and generalizability of DeePTB model across CNTs with different diameters.
To model the metal contacts, we employed a doping contact approach by setting $\rho^{\text{fix}}(r)$ in the Poisson equation (Eq.~\ref{eq:poi}).  This method, previously employed in theoretical simulations~\cite{franklin2012sub,xu2021can}, induces a built-in electrostatic field between the doped regions and channel, mimicking the band bending at the Pd electrode-CNT interface in Ohmic contacted CNT-FET~\cite{javey2003ballistic,leonard2002multiple}. We tested doping concentrations ranging from $5.33\times 10^{8}  \text{m}^{-1}$ to $1.25\times 10^{9}  \text{m}^{-1}$, with little change in ON current and subthreshold swing (SS) (see SM Sec. S7(a)). Based on these tests, we chose a doping concentration of $8.89\times 10^{8} \text{m}^{-1}$ for subsequent simulations.
Additionally, since the bottom gate covers the channel as well as the source and drain contacts in the LBG device, we set the total gate length as
$L_\text{tot}=L_\text{ch}+2\times L_\text{ext}$, including the channel length $L_\text{ch}$ and source/drain extensions $L_\text{ext}$\cite{franklin2012sub}.
Tests with different $L_\text{ext}$ values (Fig.~S6(b)) indicated that $L_\text{ext}=2.5$ nm is sufficient to screen the impact from the channel.

Figure~\ref{fig:CNT_all}(a) shows the simulated transfer characteristics for 9-nm LBG CNTs with varying diameters, demonstrating excellent agreement with experimental measurements.
For clear comparison, the gate work functions of all devices were aligned to have the same ON current $I_{\text{on}}\approx 1\times10^{-5}$ A at $V_\text{gs}=-1.0$ V.
Notably, these CNTs with varying diameters (ranging from 1,620 to 2,754 atoms) already exceed the computational limits of conventional DFT-NEGF-Poisson methods.
As illustrated in Fig.~\ref{fig:CNT_all}(a), the transfer characteristics for the four tubes are similar in the ON state but differ dramatically in the subthreshold region.
These differences arise from the distinct density of states (DOS) and band gaps of CNTs with varying diameters, resulting in different barrier heights $\Phi_\text{b}$ between source and channel in CNT-FET at OFF state.
The intrinsic DOS of these CNTs, shown in Fig.~S11, exhibit different distributions around the bias window, directly influencing the transport properties of the corresponding devices. 
In CNT-FETs, the barrier height $\Phi_\text{b}$ is illustrated by the local density of states (LDOS) as shown in Fig.~\ref{fig:CNT_all}(d)-(f). For the same p-doped device, $\Phi_\text{b}$ increases as the gate voltage $V_\text{gs}$ becomes more positive, hindering the injection of holes from the source and suppressing $I_\text{ds}$. With diameter reduction, the OFF current $I_{\text{off}}$ decreases because larger band gaps in thinner CNTs would induce higher potential barriers $\Phi_\text{b}$ in the OFF state~\cite{guo2004numerical,pal2024three}, suppressing source-drain tunneling—the major contribution to $I_{\text{off}}$.
LDOS analysis reveals a significant decrease in $\Phi_b$ with increasing CNT diameter. As shown in Fig.~S8, $\Phi_b$ in OFF states reduces from 0.48 eV for CNT(10,0) to 0.19 eV for CNT(17,0), consistent with the observed increases in $I_{\text{off}}$. 
Despite these diameter-dependent differences in OFF current, $I_{\text{on}}$ remains stable at approximately $1\times10^{-5}$ A, attributed to similar transmission functions inside the drain-source bias window in the ON state across devices with different CNT diameters (see Fig.~S7(b)). This comprehensive study of CNT diameter effects demonstrates the importance of accurately matching the simulated CNT size to experimental conditions for meaningful transfer characteristics and figures of merit such as $I_{\text{on}}/I_{\text{off}}$.

To further validate our approach against experimental data, we extended our simulations to LBG CNT-FETs with longer channel lengths (18 nm and 41 nm) using CNT(16,0) and CNT(17,0), which closely match the experimental tube diameter.
For these larger systems, the number of atoms ranges from 3,904 (CNT(16,0) with $L_\text{ch}=18$ nm) to 7,820 (CNT(17,0) with $L_\text{ch}=41$ nm), all exceeding the practical limits of conventional DFT-NEGF SCF calculations.
The simulated results, shown in Fig.~\ref{fig:CNT_all}(b) and (c), indicate that for both $L_\text{ch}=18$ and $41$ nm, the SS
agrees well with experimental data. In both cases, $I_\text{off}$ falls within the range predicted for CNT(16,0) and CNT(17,0), consistent with the experimental diameter lying between these two structures. 
The minor deviations between simulation and experiment, particularly in ON-state current and switching behavior at longer channel lengths, might be attributed to factors beyond ballistic transport, such as structural defects and scattering, which are not captured in our current model. Nevertheless, the overall agreement in key figures of merit demonstrates the validity of our approach for modeling realistic CNT-FETs.


To assess the scaling limit of LBG CNT-FETs, we further simulated devices with ultra-short channels as detailed in SM Sec.~S16.
As shown in Fig.~S10,  decreasing $L_\text{ch}$ from 9nm to 4nm leads to a sharp increase in SS from 98.88 mV/decade to 257.27 mV/decade, with $I_{\text{on}}/I_{\text{off}}$ dropping from $1\times10^5$ to $4\times10^3$.
This rapid performance degradation signals the onset of severe short-channel effects due to reduced gate electrostatic efficiency~\cite{cao2023future}, indicating that the scaling potential of the LBG geometry is limited and further device geometry improvements are necessary for continued downscaling. 

These results demonstrate the high fidelity of DeePTB-NEGF in quantum transport simulations with gate controlling, achieving DFT-level accuracy while maintaining the efficiency of TB method. 
By capturing essential physics across a wide range of nanotube diameters and channel lengths beyond the capability of conventional DFT-NEGF, this framework demonstrates excellent agreement with experimental data.
This capability enables first-principles quantum transport analysis in realistic semiconductor devices, paving the way for more precise modeling and design of next-generation nanoelectronic technologies.

\section*{Summary and Discussion}
In this work, we have developed DeePTB-NEGF, a deep learning-accelerated framework for first-principles quantum transport simulations in nanoelectronics. By integrating deep learning-based Hamiltonian prediction with the NEGF method, our approach enables efficient quantum transport calculations with or without self-consistent electrostatic effects. The framework demonstrates remarkable computational efficiency, handling break junction processes with over $10^4$ snapshots, simulating the zero-bias transmission through a 180 nm CNT ($\sim 3\times 10^4$ atoms), and modeling CNT-FETs with gate control and finite bias for channel lengths up to 41 nm ($\sim$8000 atoms)—scales previously inaccessible to first-principles methods.

For break junction systems, our framework accurately reproduced the statistical nature of experimental measurements by efficiently simulating thousands of configurations. The predicted conductance histograms captured key experimental features, including quantized conductance peaks in gold contacts and characteristic molecular junction signatures. This capability enables meaningful statistical analysis of break junction experiments with first-principles accuracy. 
For CNT-FETs, the DeePTB-NEGF-Poisson implementation successfully simulated devices with various diameters and channel lengths under finite bias conditions. The excellent agreement with experimental transfer characteristics validates our approach and demonstrates the importance of simulating devices at experimental dimensions. Our framework's ability to predict the scaling behavior of CNT-FETs provides valuable guidance for future device designs.

With these achievements, the demonstrated performance of DeePTB-NEGF establishes a new paradigm for high-throughput and large-scale quantum transport simulations in nanoelectronics. Its dual capability in simulating break junction processes and transistors at experimental size opens new opportunities for advancing molecular electronics research and semiconductor device engineering. Future developments could incorporate electron-phonon scattering, defect-induced transport, and other quantum effects beyond ballistic transport.
The framework could also be extended to more complex material interfaces and heterostructures relevant to next-generation devices.
Furthermore, recent research indicates that massively parallel GPU-accelerated NEGF implementation can significantly reduce the computation time~\cite{sawant2025eleqtronex} , offering a promising path to further optimize our framework.
By bridging the gap between fundamental physics and device-scale simulations, DeePTB-NEGF opens new possibilities for computational nanoelectronics in both research and industrial applications.

\section*{Methods}

\subsection*{Datasets}
\noindent\textbf{Metallic contacts}: The training dataset consists of 122 configurations randomly sampled from 4 independent trajectories at elongation speed $v=5\text{m/s}$.  The validation dataset includes 63 configurations from a single elongation process. Each configuration contains 304 atoms within the simulation box, with 100 atoms in both the left and right electrode regions. All configurations were generated via MD simulations at 150K in the NVT ensemble, using the LAMMPS package~\cite{lammps1995} based on a previously trained DeePMD model~\cite{andolina2021robust}. 
For DFT-NEGF calculations, we employed the TranSIESTA  code~\cite{Brandbyge2002,papior2017improvements} with generalized gradient approximation (GGA)
within the Perdew-Burke-Ernzerhof (PBE) formulation
as exchange-correlation functional and single-zeta plus
polarization(SZP) basis. Only the $\Gamma$ point was considered in reciprocal space, and norm-conserving pseudopotentials were applied. The iteration convergence threshold for the density matrix was set to $10^{-9}$.

\bigskip
\noindent\textbf{Single-molecule junctions}: The training dataset consists of 30, 60, 90, 123, 188, 208, 268 configurations randomly sampled from 14 independent trajectories at elongation speed $v=\pm2\text{m/s}$. The validation dataset includes 30 configurations from one elongation process. Each configuration contains 478 atoms within the simulation box, with 144 atoms in both the left and right electrode regions.  MD simulation was performed at 300K using the ReaxFF reactive force field~\cite{jarvi2011development}, which can efficiently describe chemical bonding through its empirical bond-order formalism~\cite{senftle2016reaxff}. Other MD simulation and DFT-NEGF parameters are the same as those used for metallic contacts. 

\bigskip
\noindent\textbf{CNT}: The training dataset comprises the band structure of zigzag CNT unit cell with different diameters (CNT(7,0), CNT(10,0), CNT(13,0), CNT(16,0)). The validation set is the band of CNT(17,0). The DFT-calculated and DeePTB-predicted bands are illustrated in SM Fig.~S3. The DFT calculation is performed in SIESTA~\cite{Soler2002} with SZP basis. The exchange-correlation functional is treated at GGA level within the PBE formulation.

\subsection*{Training Settings}
\noindent\textbf{Metallic contacts DeePTB-E3 model}:  To include the long-range interaction, the Localized Equivariant Message-passing (LEM) method is adopted in DeePTB-E3~\cite{zhouyin2025learning}, using the irreducible representations (irreps for short) of the SO(3) group as the internal equivariant features. The model consists of 4 layers of message-passing neural networks with residual updates. To represent the interactions across (off-)diagonal blocks, we use the irreps setting $32\times0e+32\times1o+32\times2e+32\times3o+32\times4e$. The latent scalar channel dimension is fixed as 64. The cutoff radius of gold atoms is set to 7.4\AA, matching the maximum radius of gold atomic orbitals in DFT-NEGF. The loss function used is 'hamil$\_$abs', which is composed of the first and second-order norms of the residuals between the target and predicted Hamiltonian/overlap matrices.
The Adam optimizer is employed with an initial learning rate of 0.01.  To accelerate convergence, the ReduceLROnPlateau learning rate schedular is employed, which decays the learning rate once the averaged loss hasn't decreased for the closest 60 epoch. We use the decay factor 0.8 in all training.

\bigskip
\noindent\textbf{Single-molecule junctions DeePTB-E3 model}:
The cutoff radius for the H, C, S, Au are 3.8, 5.6, 7.1, 6.9\AA.  We used the loss function ``hamil$\_$blas".
Unlike normal MAE or RMSE metrics that treat each matrix element equally, as in the loss function ``hamil$\_$abs", the ``hamil$\_$blas" uses a scattering mechanism to group the on(off) diagonal blocks by their atomic and bond species, computing the loss and averaged specifically. The resulting loss gives equal weights to every chemical element and bond species, helping to resolve the data scarcity when systems contain a type-unbalanced number of atoms and bonds.
The Adam optimizer is employed with an initial learning rate of 0.001.  The other training settings are same as that in metallic contacts DeePTB-E3 model.

\bigskip
\noindent\textbf{CNT DeePTB-SK model}: The fitting parameters for the empirical TB terms were represented by single neurons employing the ``strain" onsite mode with the smooth cutoff $r_s=2.5$ and decay factor $w=0.3$. The ``ploy2exp" method is used for  hopping formula with the smooth cutoff $r_s=4.0$ and decay factor $w=0.2$. For the environment descriptors in the embedding  network, we utilized se2 method\cite{zhangDeep2018} with a ResNet neural network architecture of [10, 20, 60]. In embedding, the soft and hard cutoff radius $r_s$ and $r_c$ are 4.0 \AA\quad and 6.1 \AA, respectively.
The fitting network for both hopping and onsite terms was implemented using a fully connected network with a size of [50, 50, 50]. Throughout all networks, the Tanh activation function was applied.

\section*{Data availability}

All data supporting the findings of this study are available from the corresponding author upon reasonable request.

\section*{Code availability}
The DeePTB-NEGF framework combines two open-source packages: the quantum transport code dpnegf available at \href{https://github.com/DeePTB-Lab/dpnegf}{https://github.com/DeePTB-Lab/dpnegf} and the deep learning tight-binding Hamiltonian prediction tool DeePTB accessible at \href{https://github.com/deepmodeling/DeePTB}{https://github.com/deepmodeling/DeePTB}.

\section*{Acknowledgments}
We appreciate the insightful discussions with Haoyang Pan, Daye Zheng, and Weiqing Zhou. Q. G. acknowledges the funding support from AI for Science Institute, Beijing (AISI). S. H. acknowledges the funding support from the National Key R$\&$D Program of China (2024YFA1208203). The computational resource is provided by Bohrium Cloud Platform from DP technology.


%

\end{document}